\documentclass[12pt]{article}
\usepackage{graphicx}
\usepackage{tikz}
\usepackage{siunitx}
\usepackage{textcomp}
\usepackage{amsmath}
\usepackage{float}

\usetikzlibrary{arrows,positioning} 
\tikzset{
    >=stealth',
    punkt/.style={
           rectangle,
           rounded corners,
           draw=black, very thick,
           text width=6.5em,
           minimum height=2em,
           text centered},
    pil/.style={
           ->,
           thick,
           shorten <=2pt,
           shorten >=2pt,}
}

\def\Title#1{\begin{center} {\Large {\bf #1} } \end{center}}

\begin{document}
\renewcommand*{\thefootnote}{\fnsymbol{footnote}}

\Title{Design and Simulation of the IsoDAR RFQ Direct Injection System and Spiral Inflector\footnotemark{}}
\footnotetext{Poster presented at the APS Division of Particles and Fields Meeting (DPF 2017), July 31-August 4, 2017, Fermilab. C170731}

\bigskip\bigskip

\begin{raggedright}  

{\it Philip Weigel \footnotemark{} on behalf of the DAE$\delta$LUS Collaboration\\
Laboratory for Nuclear Science\\
Massachusetts Institute of Technology\\
Cambridge, MA 02139}
\footnotetext{ plw38@drexel.edu}
\bigskip\bigskip
\end{raggedright}

\begin{abstract}

In this paper we present the development of a simulation code capable of optimizing the geometry of a spiral inflector designed for axial injection into a cyclotron. To do this, an electric field map of the device is generated by utilizing a boundary elements method and then used to track one or more particles. The information from the trajectories is then used to shorten the electrodes of the spiral inflector to adjust for fringing electric fields such that the particles end up on the mid-plane of the cyclotron. This method was also used to analyze the effects of a modified electrode geometry that can focus the beam as it travels through the device. Compared against commercial multiphysics software, the developed code produced similar results within a negligible margin of error attributed to differences in meshing and particle tracking algorithms.

\end{abstract}

\section{Introduction}

The IsoDAR (Isotope Decay-At-Rest) experiment aims to explore physics beyond the standard model by searching for anomalous neutrino oscillations indicative of sterile neutrinos \cite{PhysRevLett.109.141802}. 
The experiment requires a primary beam of $\textrm{H}_{2}^{+}$ ions to be accelerated to an energy of 60 MeV/amu at a current of 5 mA, which will be accomplished by using a high-power compact cyclotron \cite{doi:10.1063/1.4802375}. 
One of the challenges in this scheme is the injection into the cyclotron, where space charge forces are strong. 
We aim to achieve this by using a Radio-Frequency Quadrupole (RFQ) injector, which is capable of efficiently transporting and bunching the beam \cite{doi:10.1063/1.4935753}. 
Careful design and simulation of this process is necessary. 
To this end, electric and magnetic field maps are loaded into the accelerator simulation code OPAL to run start-to-end simulations of the injection system and the first few turns in the cyclotron. 
Several iterations between RFQ design and spiral inflector simulations will be necessary to optimize the geometry of the spiral inflector, i.e. to minimize the number of ions lost during injection and maximize beam current. In this paper, we will explore the shortcomings of the theoretical formulation of the spiral inflector and present methods for simulating and optimizing the electrode geometry to minimize divergence and beam losses.

\subsection{The Spiral Inflector}

\begin{figure}
	\centering
	\includegraphics[width=2in]{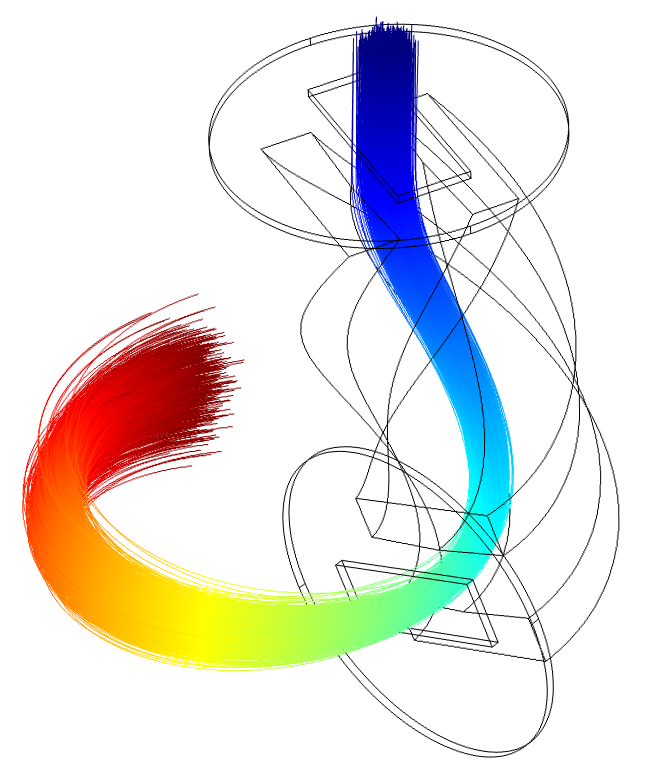}
	\caption{A model of a spiral inflector and trajectories of 70 keV $\textrm{H}_{2}^{+}$ ions. The twisted electrodes create an electric field that bends the beam onto the mid-plane of the cyclotron.}
	\label{figure:spiralinflector}
\end{figure}

For axial injection into a cyclotron, an electrostatic device is used to bend an incoming beam by $90$\textdegree \hspace{0.1em} onto the mid-plane of the cyclotron for subsequent acceleration. Different types of devices exist such as an electrostatic mirror and various inflector devices \cite{Heikkinen:1992ma}. For a centered injection scheme with low losses, a spiral inflector becomes the most favourable option, whereas hyperbolic and parabolic inflectors have entrances offset from the center of the cyclotron and electrostatic mirrors can suffer from low admittance \cite{pandit2002study}. The geometry of a spiral inflector consists of two twisted electrodes which form a strong electric field capable of bending the path of a beam of ions in conjunction with the magnetic field of the cyclotron.

\subsection{Theory}

The idea of the spiral inflector is to create an electric field that is always perpendicular to the momentum of the injected particles. This is to ensure that the particles travel along an equipotential curve onto the mid-plane of a cyclotron. For a constant magnetic field, the equations of motion for this trajectory were solved analytically in \cite{belmont1966study, toprek2000theory} to yield the parametric expressions for the particle coordinates:

\begin{equation}
    x(b) = \frac{A}{2}\Bigg[ \frac{2}{1 - 4 K^2} + \frac{\cos(2K - 1)b}{2K - 1} - \frac{\cos(2K + 1)b}{2K + 1} \Bigg]
\end{equation}

\begin{equation}
    y(b) = \frac{A}{2}\Bigg[ \frac{\sin(2K + 1)b}{2K + 1} - \frac{\sin(2K - 1)b}{2K - 1} \Bigg]
\end{equation}

\begin{equation}
    z(b) = A(1 - \sin(b))
\end{equation}

\noindent where $b$ is defined by $b = \frac{vt}{A}$ on a range of $0 \leq b \leq \frac{\pi}{2}$. These parametric equations depend on several design parameters and also the mass ($m$), velocity ($v$), and charge ($q$) of the incoming ions. From those, the following quantities are defined:

\begin{alignat*}{3}
  A = \frac{mv^2}{qE} &\qquad R_m = \frac{mv}{qB}  &\qquad K = \frac{A}{2R_m} + \frac{k^{\prime}}{2}
\end{alignat*}

The height of the spiral inflector, $A$, is determined by the magnitude of the electric field that is equal to the difference of the electrode voltages divided by the distance of the gap between them. Another parameter, K, studied more in depth by \cite{toprek2000theory}, features a free parameter $k^{\prime}$ which determines the angle at which the exit of the electrodes is tilted. The tilt parameter $k^{\prime}$ can be expressed as $k^{\prime} = \frac{\tan(\theta_t)}{\sin(\theta_t)}$ with $\theta_t$ as the physical angle of tilting with respect to the xy-plane. A consequence of the tilt is that the gap distance between the electrodes is no longer constant:

\begin{equation*}
    d(b) = \frac{d_0}{\sqrt{1 - (k^{\prime} \sin(b))^2}}
\end{equation*}

\noindent where $d_0$ is the design gap distance. Adding the tilt does not affect the height of the spiral inflector since it is defined by the value of $d_0$ and the electrode voltages, although the electric field strength will increase as $b$ increases and $d(b)$ decreases.

\subsection{Simulations}\label{section:simulations}
\subsubsection{Objective}
Many software packages have been used in the past to simulate the trajectories that ions take when traversing through the spiral inflector. With these simulations, the effects of the spiral inflector for off-centered trajectories observed in experiments can be better understood. The theoretical parametric equations that describe the equations of motion for a trajectory that is perfectly centered in the inflector and are useful for generating the geometry of the electrodes.

\subsubsection{Software}
Numerical analyses of the spiral inflector have been done in the past \cite{1289583, Ivanenko:2010zz}. For simulating space charge effects in the spiral inflector with a high number of simulated particles, OPAL-CYC has compared well with experimental results \cite{opalsim}. For preliminary designs, space charge effects and the large scale simulations are not necessary. The result of recent work has been the development of new simulation code for the purpose of lightweight simulation code to design optimized spiral inflectors. The mesh generating capabilities of gmsh \cite{geuzaine2009gmsh} are used to create surface domains for the purpose of performing an electrostatics analysis. The python BEMPP API \cite{SmigajEtAl2015} is used to run the boundary elements analysis to calculate an electric field mapping of the spiral inflector. With a magnetic field mapping of the cyclotron from another software (COMSOL, Opera, etc.) or using a constant field, particles can be tracked given initial particle parameters. The tracking algorithm used is the Boris method \cite{boris1970relativistic}, which is often used in plasma physics for tracking particles in electric and magnetic fields with high precision.

\section{Optimization}

\begin{figure}
\centering
\scalebox{0.75}{
\begin{tikzpicture}[->,node distance=1cm, auto,]
 % Nodes:
 \node[punkt] (theory) {Theory};
 \node[punkt, below=of theory] (design) {Generate Geometry};
 
 \node[below=of design] (dummy1) {};
 
 \node[punkt, left=of dummy1] (ef) {Electric Field Calculation};
 \node[punkt, below=of dummy1] (track) {Track Particles};
 \node[punkt, right=of dummy1] (adj) {Adjust Design Parameters};
 
 \draw [pil] (theory.south) to (design.north);
 \draw [pil] (design.west) to [bend right=45] (ef.north);
 \draw [pil] (ef.south) to [bend right=45] (track.west);
 \draw [pil] (track.east) to [bend right=45] (adj.south);
 \draw [pil] (adj.north) to [bend right=45] (design.east);

\end{tikzpicture}
}
\caption{Spiral Inflector Design Process}
\label{figure:designprocess}
\end{figure}
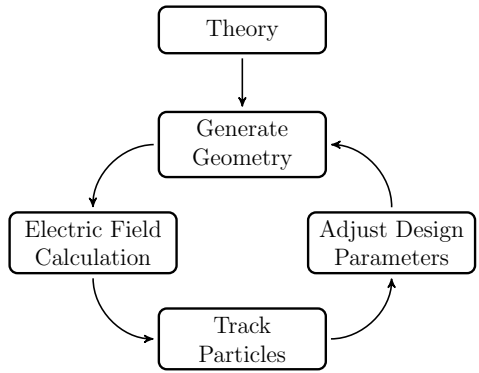

For an accelerator-based particle physics experiment such as IsoDAR, the statistics improve with the amount of beam current delivered to the target. Additionally, space charge effects for such a high current beam will become problematic and increase losses throughout the cyclotron and beam lines. For this reason, an efficient spiral inflector is a priority as it determines the shape of the beam before it begins accelerating in the cyclotron.

For the idealized spiral inflector based on the theoretical central ion trajectory, the magnitude of the electric field within the spiral inflector is assumed to be constant, or linearly increasing if there is a tilt angle, and always perpendicular to the velocity of a particle within it. The main shortcoming of these assumptions is that it neglects the effects of the fringing electric field at the entrance and exit of the spiral inflector. Essentially, the fringe fields extend the effective length of the device, which will result in the beam being inflected at an angle with respect to the mid-plane of the cyclotron. 

Using the simulations detailed in Section \ref{section:simulations}, the angle of deflection before and after the spiral inflector can be calculated and the geometry can be modified. This is typically done in a two-step process. First, an ion starting at the center of the gap at the entrance is tracked backwards. After a predetermined number of time steps, the angle of the trajectory with respect to the z-axis is measured and the spiral inflector is shortened or lengthened accordingly at the entrance by changing the minimum value of $b$ from Equations 1, 2, and 3. A similar process is then used for the exit by forward-tracking and measuring the angle between the trajectory and the mid-plane (xy-plane). For a design with no tilt or other additional features, this two step process corrects the fringe field effects. However, for designs with modifications to the electrodes or additional external surfaces that change the field strengths, multiple optimization iterations are required. 

Another way to reduce the effects of the fringe field is to introduce a set of apertures close to the entrance and exit of the device. These apertures are grounded, meaning that a particle will only be subject to the bulk of the fringe field in the space between the apertures and the electrodes. 

For the benchmarks in this paper, values relevant to the design of the IsoDAR spiral inflector are being used. $\textrm{H}_2^{+}$ ions with a kinetic energy of 70 keV are being injected from an RFQ. The voltages of the electrodes are +12.0 kV and -12.0 kV, with a gap distance of 1.8 cm and no tilt.

\begin{figure}[H]
	\centering
	\includegraphics[width=4.0in]{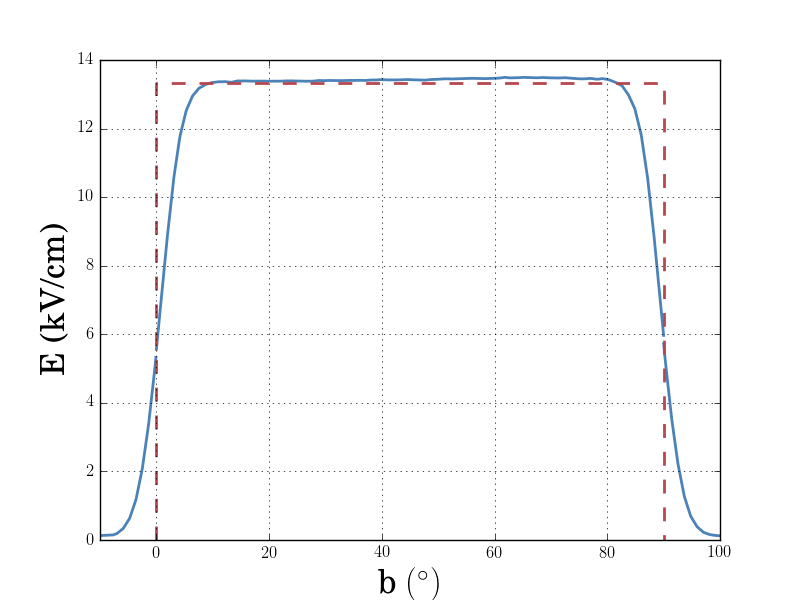}
	\caption{Analysis of the electric field magnitude over the trajectory of a simulated particle (solid) and the theoretical field magnitude (dashed). This shows that the electric field can have a fairly significant effect on the beam before it enters and after it exits the spiral inflector.}
	\label{figure:efield}
\end{figure}

\section{Benchmarking}

For the analytic treatment of the spiral inflector, the electric field is assumed to be constant which is known not to be the case. For this reason, it is important to understand how the fringe fields affect the beam that's being injected. Additionally, mapping the fringe field along the trajectory of the particle can provide insight into how the apertures should be placed. Figure \ref{figure:efield} shows the electric field magnitude over the path length of a simulated particle and the theoretical electric field magnitude for a spiral inflector with no tilt.

\begin{figure}[H]
	\centering
	\includegraphics[scale=0.45]{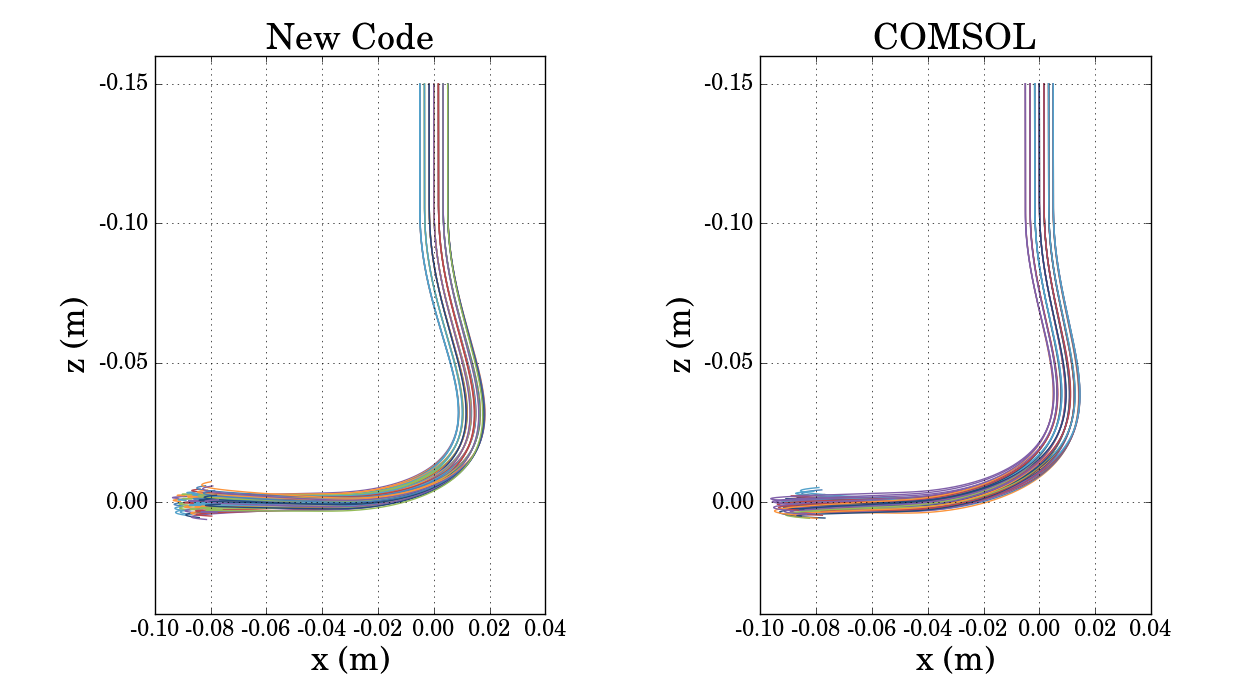}
	\caption{Left: Particle trajectories generated by a 7$\times$7 grid of 70 keV $\textrm{H}_2^{+}$ ions at z=-15 cm simulated with the newly developed code resulting in a final RMS beam size in the z-direction of 6.14 mm and z-momentum spread of 3.22\%. Right: Particle trajectories with the same parameters and spiral inflector model using COMSOL Multiphysics resulting in a final RMS beam size in the z-direction of 5.96 mm and z-momentum spread of 3.12\%.}
	\label{figure:comsol}
\end{figure}

A comparison of this newly developed simulation code with the commercial multiphysics software COMSOL yields promising results as seen in Figure \ref{figure:comsol}. This shows the difference between trajectories through the same optimized spiral inflector geometry using the newly developed code on the left and COMSOL on the right. The trajectories were generated by 7$\times$7 particles in a square grid at z = -15 cm with a side length of 1 cm centered on the z-axis. The final RMS beam sizes in the z-direction were 6.14 mm for the new code and 5.96 mm for COMSOL. Additionally, the difference between the mid-plane angles is $\Delta \theta \approx 0.8$\textdegree. Variations between the two simulations can be explained by the different meshing algorithms which result in slightly different electric fields.

Another interesting study is the analysis of a v-shaped electrode as described by \cite{Ivanenko:2010zz}. It is reported to focus the beam as it goes through the spiral inflector. The strength of this focusing effect is determined by the term $\delta$, which denotes the length of the indentation to create a chevron-shaped cross section of the electrodes. Using optimized inflector geometries, eight values of $\delta$ were used to simulate the same 7$\times$7 grid of particles as before to analyze the focusing effects.

\begin{figure}[H]
	\centering
    \includegraphics[scale=0.45]{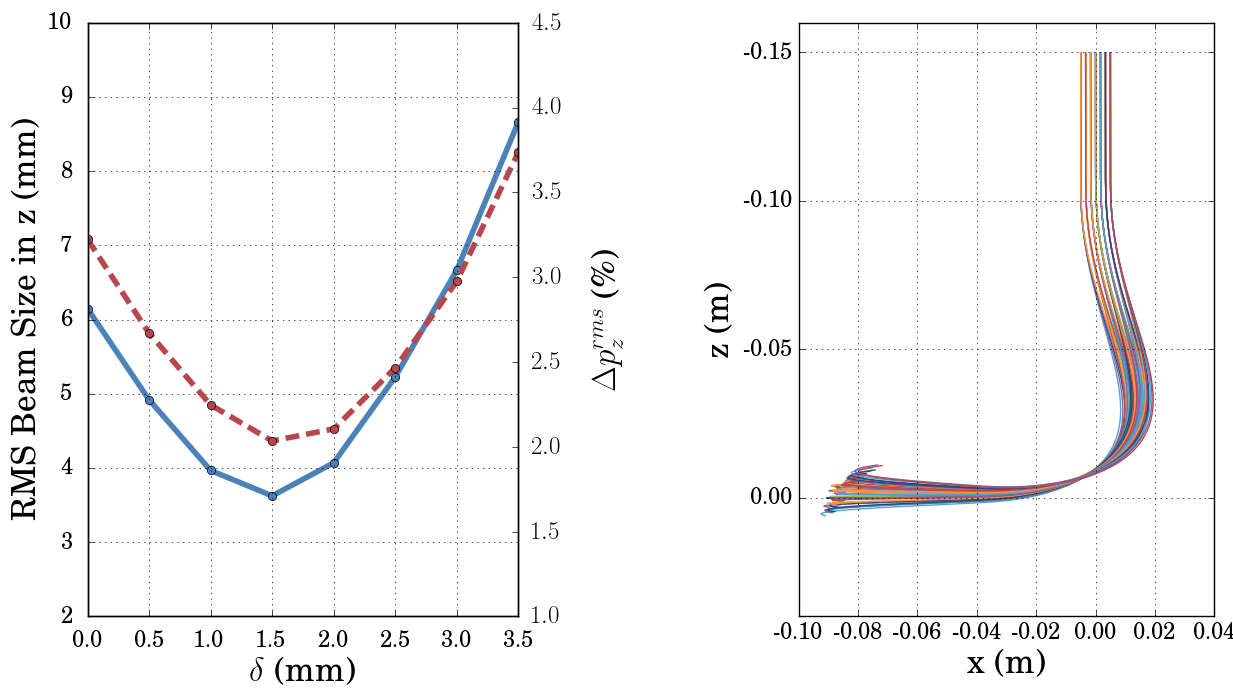}
	\caption{Left: The resulting RMS beam size in the z-direction (solid) and RMS momentum spread (dashed) from eight simulations of varying $\delta$ values. Right: Particle trajectories through a spiral inflector designed with $\delta = 3.0$ mm.}
	\label{figure:vstudy}
\end{figure}

Several simulations were run for different values of delta and both the RMS beam size and RMS momentum spread in the z-direction were analyzed and are summarized by Figure \ref{figure:vstudy}. It is clear that there is a minimum beam size and momentum spread around $\delta = 1.5$ mm of 3.62 mm and 2.03\% respectively, which then rise for higher values of $\delta$. This can be explained by an overfocusing effect caused when $\delta$ is too large, as seen by the particle trajectories displayed on the right side of Figure \ref{figure:vstudy}. By fine tuning $\delta$, it may not be necessary to add additional focusing devices after the exit of the spiral inflector which have been suggested previously to refocus the beam before being accelerated during the first turn in the cyclotron \cite{doi:10.1063/1.1435198}. For a high current beam such as the one proposed for IsoDAR, having this focusing effect may contribute to minimizing beam losses.

\section{Conclusion}

The purpose of this newly developed code is to be able to design an optimized spiral inflector by an iterative process of calculating the electric fields and tracking particles, and removing parts of the electrodes to adjust for the fringe field effects. A comparison between the code and COMSOL demonstrated the accuracy, which provides a confirmation that the simulations produce similar results to the widely used multiphysics software. For the ongoing design of the RFQ direct injection system, new optimized spiral inflector designs can now be quickly generated to account for changes necessitated by RFQ simulations (e.g. changes in incoming particle energy). Looking forward, work will be done on improving the results of the optimization mechanism and the inclusion of basic space charge calculations to be able to adjust the focusing in the presence of self-fields.

\end{document}